\documentclass[11pt]{article}
\usepackage[pdftex]{graphicx} 
\usepackage[comma,numbers,sort&compress]{natbib} 
\usepackage{amsfonts,amsmath,amssymb,amsthm,mathtools} 
\usepackage{wrapfig}  
\usepackage{mdwlist}  
\usepackage{sidecap}  
\usepackage{bbm}
\usepackage{bm,amsbsy} 
\usepackage{algorithm,algorithmic}

\newcommand{\vx}{\mathbf{x}}

\newcommand{\Dat}{\mathcal{D}}
\newcommand{\trp}{{^\top}} 
\newcommand{\inv}{^{-1}}
\newcommand{\Nrm}{\mathcal{N}}

\newcommand{\vz}{\mathbf{z}}

\newcommand{\vs}{\mathbf{s}}

\newcommand{\vtheta}{\mathbf{\ensuremath{\bm{\theta}}}}

\newcommand{\vy}{\mathbf{y}}

\begin{document}


\title{A note on privacy preserving iteratively reweighted least squares}
\author{Mijung Park and Max Welling \\
QUvA Lab. University of Amsterdam \\
\texttt{\{mijungi.p,  welling.max\}@gmail.com}}
 \date{}

\maketitle

\vspace{-0.5cm}

\centerline{\textbf{Abstract}}
\hbox to \textwidth{\hrulefill} 

Iteratively reweighted least squares (IRLS) is a widely-used method in machine learning to estimate the parameters in the generalised linear models. In particular, IRLS for L1 minimisation under the linear model provides a closed-form solution in each step, which is a simple multiplication between the inverse of the weighted second moment matrix and the weighted first moment vector.
When dealing with privacy sensitive data, however, developing a privacy preserving IRLS algorithm faces two challenges. First, due to the inversion of the second moment matrix, the usual sensitivity analysis in differential privacy incorporating a single datapoint perturbation gets complicated and often requires unrealistic assumptions. 
Second, due to its iterative nature, a significant cumulative privacy loss occurs. However,  adding a high level of noise to compensate for the privacy loss hinders from getting accurate estimates.    
Here, we develop a practical algorithm that overcomes these challenges and outputs privatised and accurate IRLS solutions. 
In our method, we analyse the sensitivity of each moments separately and treat the matrix inversion and multiplication as a post-processing step, which simplifies the sensitivity analysis. 
Furthermore, we apply the {\it{concentrated differential privacy}} formalism, a more relaxed version of differential privacy, which requires adding a significantly less amount of noise for the same level of privacy guarantee, compared to the conventional and advanced compositions of differentially private mechanisms.  

\hbox to \textwidth{\hrulefill}

\vspace{-0.5cm}

\section{Introduction}

Differential privacy (DP) algorithms provide strong privacy guarantees by typically perturbing some statistics of a given dataset, which appear in the outputs of an algorithm \cite{Dwork14}.  The amount of noise added for the perturbation is set in order to compensate for any difference in the probability of any outcome of an algorithm by adding a single individual's datapoint to or removing it from the data.  So, in order to develop a DP algorithm, one first needs to analyse the maximum difference in the probability of any outcome, which often called {\it{sensitivity}}, to set the level of additive noise.

In this note, we're interested in developing a privacy preserving iteratively reweighted least squares (IRLS) method. In the compressed sensing literature \cite{4518498}, IRLS is used for solving the L-1 minimisation problem, in which a closed-form update of parameters in each step is available. This IRLS solution in each step is a simple multiplication between the inverse of the weighted second moment matrix and the weighted first moment vector. Due to the inverse of the second moment matrix, analysing the sensitivity becomes challenging. Previous work \cite{Sheffet15} assumes each feature of each datapoint is $i.i.d.$ drawn from a standard normal distribution, and analysed the sensitivity of the inverse of the second moment matrix. 
Unfortunately, the assumption on each features being independent is often not realistic.  

%

Another challenge in developing a privacy preserving IRLS method comes from the iterative nature of the IRLS algorithm. The conventional DP composition theorem (Theorem 3.16 in \cite{Dwork14}) states that multiple iterations of a $\epsilon$-DP algorithm faces  a linearly degrading privacy, which yields $J\epsilon$-DP after $J$ iterations. A more advanced composition theorem (Theorem 3.20 in \cite{Dwork14}) yields ($\sqrt{2J\log(1/\delta)}\epsilon + J \epsilon(e^{\epsilon}-1)$, $\delta$)-DP. 
The new variable $\delta$ (stating the mechanism's failure probability) that needs to be set to a very small value, which makes the cumulative privacy loss still relatively high. To compensate for the privacy loss, one needs to add a significant amount of noise to the IRLS solution to avoid revealing any individual information from the output of the algorithm.

In this note, we tackle these challenges by : (1) we analyse the sensitivity of the weighted second moment matrix and the weighted first moment vector separately and perturb each moment by adding noise consistent with its own sensitivity. Then, we do the multiplication of the inverse of perturbed second moment matrix and the perturbed first moment vector. This inversion and  multiplication can be viewed as a post-processing step, which doesn't alter the privacy level. Since we perturb each moment separately, this method does not require any restrictive assumptions on the data. In addition, the noise variance naturally scales with the amount of data. (2) we apply the {\it{concentrated differential privacy}} formalism, a more relaxed version of differential privacy, to obtain more accurate estimates for the same cumulative privacy loss, compared to DP and its ($\epsilon, \delta$)-relaxation. In the following, we start by describing our privacy preserving IRLS algorithm. 



\section{Privacy preserving IRLS}

Given a dataset which consists of $N$ input-output pairs $\{ \vx_i, y_i\}_{i=1}^N$ where we assume $ ||\vx_i||_2 \leq 1$ and $||y_i|| \leq 1$. 
The iteratively reweighted least squares solution has the form:
\begin{eqnarray}
\hat{\vtheta}^{(t)}_{irls} &=& (X \trp S X)^{-1} X\trp S \vy : =  B \inv A
\end{eqnarray} where $X \in \mathbb{R}^{N \times d}$ is a design matrix in which the $i$th row is the transposed $i$th input $\vx\trp$ (of length $d$), and $\vy$ is a column vector of outputs. We denote $B = \frac{1}{N}X \trp S X$, and $A = \frac{1}{N} X\trp S \vy$. Here $S$ is a diagonal matrix with diagonal $\vs = |\vy - X \hat{\vtheta}^{(t-1)}|^{p-2}$. Here we set $p=1$ and compute $L1$ norm constrained least squares. 
To avoid dividing by $0$, we set 
\begin{equation}
\vs_i = \frac{1}{\max(1/\delta, \; |\vy_i - X_i \hat{\vtheta}^{(t-1)}|)}, 
\end{equation} where $X_i$ is the $i$th row. $\delta$ sets the sparsity (number of non-zero values) of the IRLS solution. 

We will perturb each of these statistics $A$ and $B$ by certain amounts, such that each statistic is $\epsilon'-$differentially private in each iteration. 

\paragraph{$\epsilon'$-differentially private moment $A$ by Laplace mechanism.}

For perturbing $A$, we use the Laplace mechanism. To use the Laplace mechanism, 
we first need to analyse the following L1-sensitivity:
\begin{eqnarray}\label{L1sen}
\Delta A&:=& \max_{\Dat, \tilde{\Dat} \in \mathbb{N}^{|\chi|}, \; ||\Dat-\tilde{\Dat}||_1 =1} || \tfrac{1}{N} X \trp S \vy - \tfrac{1}{N}\tilde{X} \trp \tilde{S} \tilde{\vy} ||_1
= \max_{\vx_k, \tilde{\vx}_k} ||\frac{1}{N} \vx_k s_k y_k\trp -  \frac{1}{N} \tilde{\vx}_k \tilde{s}_k \tilde{y}_k\trp  ||_1, \nonumber \\
%
%
&\leq&\frac{1}{N} \sum_{l=1}^d | \vx_{k, l}s_k y_{k} | + \frac{1}{N} \sum_{l=1}^d | \tilde\vx_{k, l}\tilde{s}_k \tilde{y}_{k} |, \mbox{triangle inequality} \nonumber \\
%
&\leq& \frac{s_k}{N} \sum_{l=1}^d | \vx_{k, l}| +    \frac{\tilde{s}_k}{N} \sum_{l=1}^d | \tilde\vx_{k, l}  |,  \; \mbox{since $|y_{k'}| \leq 1 $ and $ |\tilde{y}_k| \leq 1$}, \nonumber \\
&\leq&  \frac{2\delta\sqrt{d}}{N}, \; \mbox{since $s_k \leq \delta $ and $|\vx_l|_1\leq \sqrt{d}$}.
\end{eqnarray}

Hence, the following Laplace mechanism  produces $\epsilon'$-differentially private moment of $A$:
\begin{equation}\label{Laplace}
\tilde{A} = A + (Y_1, \cdots, Y_d),
\end{equation} where $Y_i \sim^{i.i.d.} \mbox{Laplace}(\frac{2\delta \sqrt{d}}{N\epsilon'})$. 

\paragraph{$\epsilon'$-differentially private moment $A$ by Gaussian mechanism.}
One could perturb the first moment by using the Gaussian mechanism. To use the Gaussian mechanism, one needs to analyse the L2-sensitivity, which is $\Delta_2 A = 2\delta/N$ straightforwardly coming from  Eq.\eqref{L1sen}
\begin{equation}\label{Gauss}
\tilde{A} = A + (Y_1, \cdots, Y_d),
\end{equation} where $Y_i \sim^{i.i.d.} \mbox{Gaussian}(0, \sigma^2)$, where $\sigma \geq c \Delta_2 A / \epsilon'$ for $c^2 \geq 2 \log(1.25/\delta)$.

\paragraph{$\epsilon'$-differentially private moment $B$.}
We perturb $B$ by adding Wishart noise following \cite{SYM_Mat_PERT}, which provides strong privacy guarantees and significantly higher utility than other methods (e.g., \cite{DworkTT014, NIPS2014_5326, NIPS2012_4565}) as illustrated in \cite{SYM_Mat_PERT} when perturbing positive definite matrices.

To draw Wishart noise, we first draw Gaussian random variables: 
\begin{eqnarray}
\vz_i \sim \Nrm\left(0, \frac{\delta}{2\epsilon' N} I_{d}\right), \mbox{ for } i=\{1, \cdots, d+1\}, 
\end{eqnarray} and construct a matrix $Z := [\vz_1, \cdots, \vz_{d+1}] \in \mathbb{R}^{d \times (d+1)}$
\begin{equation}\label{Wishart}
\tilde{B} := B + Z Z\trp
\end{equation} then $\tilde{B}$ is a $\epsilon'$-differentially private second moment matrix. 
Proof follows the paper  \cite{SYM_Mat_PERT}. The matrix $ZZ\trp$ is a sample from a Wishart distribution $\mbox{W}(ZZ\trp|\frac{\delta}{2\epsilon' N} I_{d}, d+1)$.
The probability ratio between a  noised-up version  $\tilde{B}$ given a dataset $\Dat$ (where $B$ is the exact second moment matrix given $\Dat$) and given a neighbouring dataset $\Dat'$ (where $B'$ is the exact second moment matrix given $\Dat'$) is given by    
\begin{eqnarray}
\frac{\mbox{W}(\tilde{B} - B|\frac{\delta}{2 \epsilon' N} I_{d}, d+1) }{\mbox{W}(\tilde{B} - B'|\frac{\delta}{2 \epsilon' N} I_{d}, d+1)} &=& \frac{\exp(- \frac{\epsilon' N}{\delta} \mbox{tr}(\tilde{B} - B))}{\exp(- \frac{\epsilon' N}{\delta} \mbox{tr}(\tilde{B} - B'))}, \\
&=& \exp(\frac{\epsilon' N}{\delta} \mbox{tr}( B- B')), \\
&=& \exp(\frac{\epsilon' N}{\delta} \frac{1}{N}\mbox{tr}( s_k\vx_k\vx_k\trp- \tilde{s}_k\tilde{\vx}_k\tilde{\vx}_k\trp)), \\
&=&  \exp(\frac{\epsilon'}{\delta}(s_k\vx_k\trp\vx_k - \tilde{s}_k\tilde{\vx}_k\trp\tilde{\vx}_k)), \\
&\leq&\exp(\epsilon'), \; \mbox{since $0\leq\vx_k\trp\vx_k \leq1$, and $0\leq s_k\leq \delta$}.\nonumber
\end{eqnarray}


\section{Concentrated differential privacy for IRLS}

Here we adopt a relaxed version of DP, the so-called concentrated differential privacy (CDP) in order to significantly lower the amounts of noise to add to the moments without compromising on cumulative privacy loss over several iterations. 

According to Theorem 3.5 in \cite{2016arXiv160301887D}, any $\epsilon$-DP algorithm is ($\epsilon(\exp(\epsilon)-1)/2, \epsilon$)-CDP. Furthermore, theorem 3.4 states that J-composition of ($\mu, \tau$)-CDP mechanism guarantees ($\sum_{i=1}^J \mu_i, \sqrt{\sum_{i=1}^J \tau_i^2}$)-CDP. Suppose we perturb some key statistic in each IRLS iteration using the Laplace mechanism. Denote the difference in statistic given dataset x and y by $\Delta S := S(x)- S(y)$. The conventional composition theorem says that I should add Lap($\Delta S J/\epsilon$) in each iteration to ensure $\epsilon$-DP after J iterations.  Now suppose we perturb the key statistic in each iteration by adding Laplace noise drawn from Lap($\Delta S / \epsilon'$), which, according to Theorem 3.5 in \cite{2016arXiv160301887D}, gives us a ($\epsilon'(\exp(\epsilon')-1)/2, \epsilon'$)-CDP solution.  According to Theorem 3.4 in \cite{2016arXiv160301887D}, after J iterations, we obtain a ($J\epsilon'(\exp(\epsilon')-1)/2, \sqrt{J } \epsilon'$)-CDP solution. 
What we want to make sure is if the expected privacy loss is equal to our privacy budget $\epsilon$, i.e., $J \epsilon'(\exp(\epsilon') -1)/2 = \epsilon$. Using Taylor's expansion, we can rewrite the left hand side by $J \epsilon' (1+\epsilon' + \sum_{j=2}^\infty \frac{\epsilon'^j}{j!} - 1)/2 = \epsilon$, which we can lower bound by ignoring the infinite sum, $J \epsilon'^2 /2 \leq \epsilon$. Hence, the largest $\epsilon'$ should be less than equal to $\sqrt{2 \epsilon/J}$.

This says, in each iteration, the key statistic should be perturbed by adding Laplace noise drawn from Lap($\sqrt{J} \Delta S / \sqrt{2\epsilon}$), in order to obtain a  ($\epsilon, \sqrt{ 2\epsilon}$)-CDP solution after $J$ iterations.
In the IRLS algorithm, we have two statistics to perturb in each iteration. Suppose we perturb each statistic to ensure $\epsilon'$-DP. Then, we can modify the result above by replacing $J$ with $2J$ for the IRLS algorithm. Hence, each perturbation should result in $\epsilon'$-DP parameter, where 
$\epsilon' := \sqrt{\frac{2 \epsilon}{2J}} = \sqrt{\frac{\epsilon}{J}}.$
This gives us the $\epsilon$-CDP IRLS algorithm below.
\begin{algorithm}[h]
\caption{ ($\epsilon, \sqrt{ 2\epsilon}$)-CDP IRLS algorithm via moment perturbation}\label{algo:CDPIRLS}
\begin{algorithmic}
\vspace{0.2cm}
\REQUIRE Dataset $\Dat$
\vspace{0.2cm}
\ENSURE $\epsilon$-IRDP least squares solution after $J$-iteration
\vspace{0.2cm}
\STATE (1) Compute $A= \tfrac{1}{N} X\trp \vy$  and add either Laplace or Gaussian noise by Eq.\eqref{Laplace} or Eq.\eqref{Gauss} 
\vspace{0.2cm}
\STATE (2) Compute $B = \tfrac{1}{N} X\trp X$  and add Wishart noise by Eq.\eqref{Wishart}
\vspace{0.2cm}
\STATE (3) Compute the $\epsilon$-CDP least squares solution by $\vtheta_{cdpirls}:=\tilde{B}\inv \tilde{A}$.
\vspace{0.2cm}
\end{algorithmic}
\end{algorithm}

\vspace{-0.5cm}
\section{Experiments}
Our simulated dataset consists of $N$ datatpoints, each with $d$ dimensional covariates, generated using {\it{i.i.d.}} draws $\vx_i \sim \Nrm(0, I_d)$, then normalise $X$ such that the largest squared L2 norm is 1. We generated the true parameter $\vtheta \in \mathbb{R}^d$ from $\Nrm(0, I_d)$. We generated each observation $y_i$ from $\Nrm(X\vtheta, \sigma^2 I)$, where $\sigma^2 = 0.01$. We also normalised $Y$ such that the largest squared L2-norm is 1.
\begin{SCfigure}[10][h]
\centering
\includegraphics[width=0.55\textwidth]{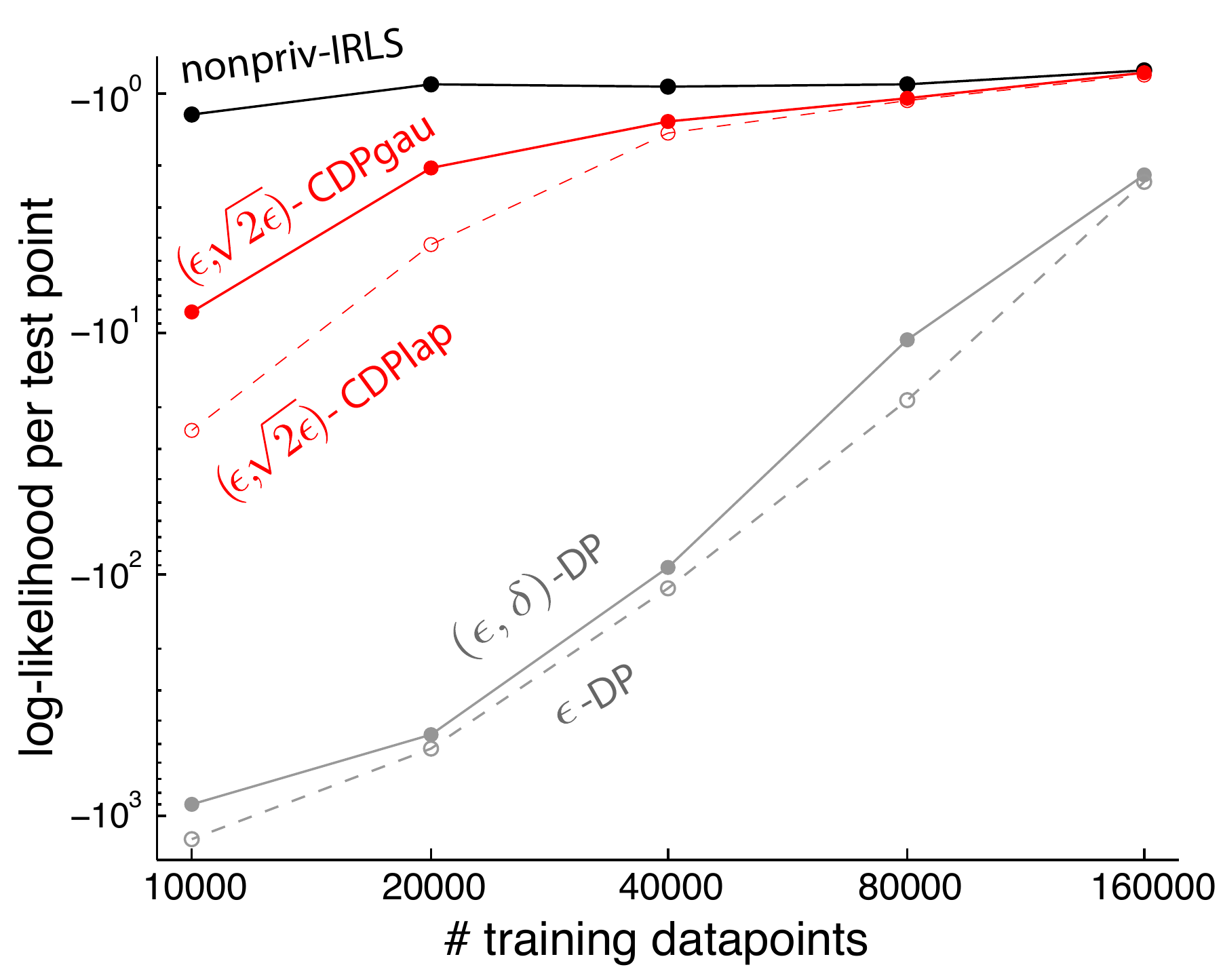}
  \caption{ We tested ($\epsilon, \sqrt{ 2\epsilon}$)-CDP-IRLS (gau: Gaussian mechanism for mean perturbation, lap: Laplace mechanism for mean perturbation),  $\epsilon$-DP-IRLS (using the conventional composition theorem), and ($\epsilon, \delta$)-DP-IRLS (using the Advanced composition theorem)  for $d=10$  and $\epsilon=0.9$ with varying $N$ for which we generated $20$ independent datasets. For each IRLS solution, we computed the log-likelihood of test data ($10 \%$ of training data), then divided by the number of test points to show the log-likelihood per test point. CDP-IRLS requires significantly less data than DP-IRLS for the same level of expected privacy.}
\label{fig:figure1}
\end{SCfigure}

  \newpage

\section*{Acknowledgements}
This work is supported by Qualcomm.

\small
\bibliographystyle{unsrt}
\bibliography{ref_CDP_IRLS.bib}

\end{document}